\documentclass{tihany}

\newcommand{\beq}{\begin{equation}}
\newcommand{\eeq}{\end{equation}}

\newcommand{\bp}{{\bbox{p}}}

\newcommand{\bq}{{\bbox{q}}}
\newcommand{\bk}{{\bbox{k}}}
\newcommand{\bK}{{\bbox{K}}}

\newcommand{\tp}{{\tilde p}}
\newcommand{\tk}{{\tilde k}}

\newcommand{\half}{{\textstyle {1\over 2}}}
\newcommand{\bbox}[1]{{\vec #1}}

\begin{document}

\title{Multi-Boson Effects in Bose-Einstein Interferometry}

\author{Ulrich Heinz}

\address{Physics Department, The Ohio State University, 174 West 18th Avenue,
Columbus, OH 43210, USA\\
E-mail: heinz@mps.ohio-state.edu}

\maketitle

\abstracts{Multi-boson symmetrization effects on two-particle 
Bose-Einstein interferometry are studied for ensembles with arbitrary 
multiplicity distributions. In the general case one finds interesting 
residual correlations which require a modified framework for extracting 
information on the source geometry from two-particle correlation 
measurements. In sources with high phase-space densities, multi-boson 
effects modify the Hanbury Brown--Twiss (HBT) radius parameters and 
simultaneously generate strong residual correlations. We clarify their 
effect on the correlation strength (intercept parameter) and thus 
explain a variety of previously reported puzzling multi-boson 
symmetrization phenomena. For event ensembles of (approximately) fixed
multiplicity, the residual correlations lead to a minimum in the
correlation function at non-zero relative momentum, which can be
practically exploited to search, in a model-independent way, for
multi-boson symmetrization effects in high-energy heavy-ion experiments.    
}

\section{Introduction}\label{sec1}

Heavy-ion collisions at high energies produce large numbers of secondary
particles. Even if these particles are created independently (``chaotic 
source'') and are free from final state interactions, the measured 
$n$-body momentum spectra still do not simply factor into a product of 
single-particle spectra. Instead, symmetrization of the many-body wave 
function with respect to the exchange of identical particles leads (in 
the case of bosons) to so-called Bose-Einstein correlations. 

Bose-Einstein correlations affect the measured cross sections whenever
more than one boson occupies an elementary phase-space cell. This is the 
basis of Bose-Einstein or intensity interferometry: for a source of size
$R$, symmetrization effects lead to an enhancement of, for example, the 
two-particle cross-section at relative momenta $q<1/R$. The theory which 
allows to extract the source size $R$ from two-particle correlation 
data assumes, in its simplest and usually employed form,\cite{reviews} 
that the effect is dominated by symmetrization with respect to an 
exchange of the two observed particles, and that the additional 
symmetrization with respect to all the other identical bosons in 
the multiparticle final state can be neglected.

This is only true if the phase-space density of the source at freeze-out 
(i.e. at point of last interaction) is sufficiently small. In regions of
high phase-space density, multi-boson symmetrization effects change the
shape even of the single-particle spectra and two-particle correlations.
They lead to a clustering a low momenta in momentum space and at low 
separation in coordinate space. For example, in an infinite medium they 
turn an exponential momentum distribution of distinguishable elementary 
pion production sources into a Bose-Einstein distribution for the observed
pion momenta.\cite{L00,diss} And in two-pion interferometry, the effective 
source extension extracted from the two-pion correlation function is 
smaller than the r.m.s. width of the spatial distribution of the individual 
pion emitters.\cite{Zh97,W98}

In addition to these generally accepted phenomena, multi-boson 
symmetrization effects also cause other, somewhat more confusing effects.
For example, it has been noted\cite{Zh97,W98} that in ensembles with 
fixed total pion multiplicity multi-pion effects reduce the intercept 
of the two-pion correlation function at vanishing relative momentum. This 
has been interpreted\cite{ZC98} as a sign of ``effective coherence'' in 
the source which becomes significant if its phase-space density is large 
and which will lead to a ``pion laser''\cite{Pratt93} once a critical 
phase-space density is exceeded. This interpretation will be shown to be 
incorrect: The intercept reduction happens even for completely chaotic 
sources without any phase coherence, and even the formation of a Bose 
condensate is not accompanied by ``pion lasing''. We found\cite{HSZ00}
that the reduced intercept reflects {\em residual correlations} among 
the pions which are not directly related to the Hanbury Brown-Twiss 
correlations which are exploited in Bose-Einstein interferometry. These
residual correlations are generic in the sense that they vanish only for 
very specific cases where the multiplicity distribution of the ensemble 
takes on a particular form.\cite{HSZ00} When present, they render the 
extraction of the source size from Bose-Einstein interferometry more 
difficult. 

\section{Multi-boson formalism}\label{sec2}

Due to space reasons, I can give here only a very rudimentary account of
the mathematics; all technical details can be found elsewhere.\cite{HSZ00} 
Following earlier work by others,\cite{ZC98,Pratt93,GKW,CH94} the source 
is parametrized by a density operator constructed out of a superposition 
of $n$-particle states emitted from classical source currents which 
themselves are Gaussian wave packets. The number of these source current 
wave packets, their centers in phase-space, and the pion multiplicity 
$n$ all have completely arbitrary (although normalizable) distributions. 
The source currents emit pions with random phases which ensures complete 
source chaoticity. For technical reasons, the $n$-pion states used in the 
construction are not normalized,\cite{ZC98,Pratt93,HSZ00,fn1} but the 
density operator is. Calculating this normalization is the only part of 
the problem which cannot be done analytically even for the simple case of 
nonrelativistic, instantaneous Gaussian wave packets, but requires the 
numerical evaluation of a recursion relation.\cite{CGZ95}

As found by Pratt\cite{Pratt93}, all $n$-particle cross sections can be
expressed in terms of the so-called ``ring integrals''
 \bea
 \label{21}
    G_i(\bp_1,\bp_2) &=& \int G_1(\bp_1,\bk_1)\,d\tk_1\,
    G_1(\bk_1,\bk_2)\cdots d\tk_{i-1} 
    G_1(\bk_{i-1},\bp_2)\,,
 \eea
where $d\tk\,{=}\,d^3k/E_k$. They describe the effect on the two-particle
exchange amplitude (with respect to the interchange of momenta $\bp_1$ and 
$\bp_2$) of the symmetrization with respect to $i{-}1$ additional particles
with momenta $\bk_1,\dots,\bk_{i-1}$. They can be generated recursively 
from the elementary two-particle exchange amplitude $G_1(\bp_1,\bp_2)$, 
the Fourier transform of the source Wigner density $g(x,K)$:
 \bea
 \label{24}
   G_1(\bp_1,\bp_2) =
   \int \! d^4x\, g\left(x,\half(p_1{+}p_2)\right)
   e^{-i(p_1{-}p_2)\cdot x}
   \equiv \int \! d^4x\, g\left(x,K\right)
   e^{-i q\cdot x} .
 \eea
For a non-relativistic spherical Gaussian Wigner density with spatial 
width $R$ and momentum spread $\Delta$, which emits particles 
instantaneously $\sim \delta(t)$, the ring integrals can be calculated
analytically:\cite{ZC98,HSZ00}
 \bea
 \label{57a}
   &&G_n(\bp_1,\bp_2) =
   {c_n \sqrt{E_1\,E_2}\over (2\pi\Delta^2)^{3\over 2}}\,
   \exp\left( - {\bK^2\over 2\Delta_n^2} 
              - {R_n^2\,\bq^{\,2}\over 2}\right),
 \\
 \label{57b}
   &&R_n^2 = a_n\,R^2\, ,\quad \Delta_n^2 = a_n\,\Delta^2\, ,
                         \quad v = 2 R \Delta \geq 1\, ,
 \\
 \label{57c}
   &&a_n = {1\over v}\, {(v{+}1)^n + (v{-}1)^n \over (v{+}1)^n - (v{-}1)^n}
   \leq 1\, ,\quad 
     c_n = \left( 2^{2n}\, v \over (v{+}1)^{2n} -(v{-}1)^{2n} 
                \right)^{3/2} \! \leq 1\, .\ \ 
 \eea
One sees that the higher order ring integrals are Gaussians with reduced
widths $R_n$ and $\Delta_n$ in both coordinate and momentum space, 
reflecting the bosonic clustering at low momenta and relative distances 
mentioned in the Introduction. The strength of these clustering effects
is controlled by the single parameter $v=2R\Delta$ which measures the
phase-space volume of the source. The above equations remain correct
for expanding Gaussian sources (after suitable redefinition\cite{HSZ00} 
of the parameters $R$ and $\Delta$), and they are easi\-ly generalized
for Gaussians with different widths along the three Cartesian directions.

\section{Bose-Einstein and residual corelations}\label{sec3}

For this very general class of models, the two-particle correlation 
function can be written down explicitly as follows: 
 \begin{eqnarray}
 \label{62}
   C_2(\bp_1,\bp_2) &=& 
     {\sum_{i,j=1}^\infty h_{i+j}\, G_i(\bp_1,\bp_1)\, G_j(\bp_2,\bp_2)
      \over
      \sum_{i,j=1}^\infty h_i\, h_j\, G_i(\bp_1,\bp_1)\, G_j(\bp_2,\bp_2)}
 \nonumber\\
   &&\times \left( 1 +
   {\sum_{i,j=1}^\infty h_{i+j}\, G_i(\bp_1,\bp_2)\, G_j(\bp_2,\bp_1)
    \over
    \sum_{i,j=1}^\infty h_{i+j}\, G_i(\bp_1,\bp_1)\, G_j(\bp_2,\bp_2)}
   \right)
 \nonumber\\
   &\equiv& C_2^{({\rm res})}(\bq,\bK) 
   \Bigl( 1 + R_2(\bq,\bK) \Bigr) \, .
 \end{eqnarray}
The sums over $i$ and $j$ take into account symmetrizations with respect
to an increasing number of other particles in the final state. The weights
$h_i$ are related to the measured pion multiplicity distribution $p_n$
by
 \beq
 \label{35}
   h_i = \sum_{n=i}^\infty p_n\, {\omega(n-i)\over\omega(n)}\, ,
 \eeq
where $\omega(n)$ is (up to trivial factors) the factor required
to normalize the $n$-particle density operator\cite{HSZ00} and must be 
obtained by numerically solving the recursion relation\cite{CGZ95}
 \beq
 \label{23}
   \omega(n) = {1\over n} \sum_{i=1}^n C_i\, \omega(n-i)\,,
 \eeq
with $\omega(0)=\omega(1)=1$ and $C_i=\int d\tp\,G_i(\bp,\bp)=c_n a_n^{3/2}$.
Obviously, the $\omega(n)$ and the weights $h_i$ depend non-trivially 
on the phase-space volume factor $v$.

\begin{figure}[t]
\begin{center}
\epsfxsize=8cm \epsfbox{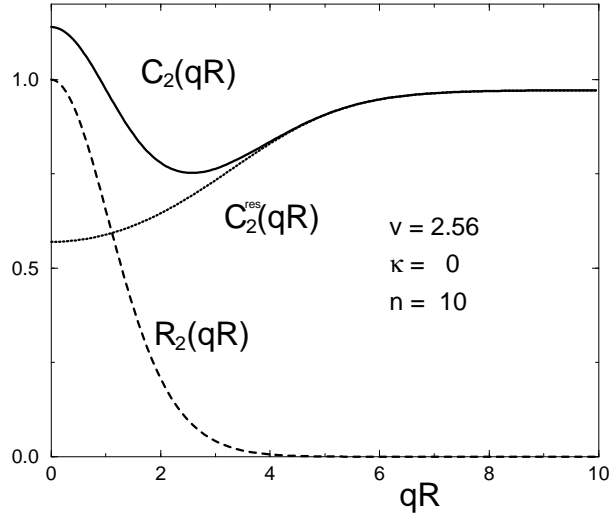}
\end{center}
\caption{Two-particle correlation function, indicating separately the
Bose-Einstein correlator $R_2$ and the residual correlations 
$C_2^{(\rm res)}$, for a fixed pion multiplicity $n{\,=\,}10$, a 
phase-space volume factor $v{\,=\,}2R\Delta{\,=\,}2.56$, and vanishing
reduced pair momentum $\kappa{\,=\,}K/\Delta{\,=\,}0$. 
\label{F1}}
\end{figure}

For dilute systems ($v\gg 1$), the higher order ring integrals can be 
neglected, and all the sums in (\ref{62}) are dominated by their fist 
terms. In this case the prefactor $C_2^{\rm (res)}$ in (\ref{62}) reduces
to a normalization constant ${\cal N}=h_2/h_1^2$, and the correlator
assumes its standard form:\cite{reviews}
 \begin{eqnarray}
 \label{63}
   C_2(\bp_1,\bp_2) = {\cal N} \left( 1 +
   {|G_1(\bp_1,\bp_2)|^2 \over G_1(\bp_1,\bp_1)\, G_1(\bp_2,\bp_2)}
   \right)\,.
 \end{eqnarray}
In general, however, the prefactor is a non-trivial function of both the 
pair momentum $\bK$ and the relative momentum $\bq$. These residual 
correlations do not involve the exchange of momenta $\bp_1$ and $\bp_2$
and thus are not related to Bose-Einstein interferometry; they cannot be 
used to extract the source radius $R$. They do, however, modulate the 
correlation function and modify its intercept at $q=0$. A typical example
is shown in Figure~\ref{F1}. One sees the genuine (Gaussian) Bose-Einstein
correlation function $R_2(\bK,\bq)$ sitting on top of the residual
correlation; the latter starts out at the value ${\cal N}{\,=\,}h_2/h_1^2$ 
at $q{\,=\,}\infty$ but decreases to a much smaller value at $q\,{=}\,0$.

The source size signal is contained in the function $R_2$ in (\ref{62}). 
Note that $R_2$ goes to 1 as $q\to 0$; thus at $q{\,=\,}0$ the exchange 
term is as large as the direct term, implying a fully chaotic source (as 
it should be by construction). The reduced intercept at $q{\,=\,}0$ thus 
has nothing to do with coherence. 

The residual correlations can be positive or negative, enhancing or 
decreasing the intercept at $q{\,=\,}0$ above or below the value 
$2{\cal N}$. Negative residual correlations appear, for example, in 
ensembles with fixed pion multiplicity or with a Poissonian multiplicity 
distribution; on the other hand, a Bose-Einstein multipliciy distribution 
can lead to large positive residual correlations.\cite{HSZ00} The 
residual correlations disappear in the dilute limit, $v\to \infty$, but 
generically become strong for large phase-space densities or small
values of $v$. At the moment there are only two special multiplicity 
distributions known\cite{HSZ00} for which residual correlations are 
completely absent, for all allowed\cite{fn2} values of $v$.

The limit $v{\,\to\,}1$ is particularly interesting. In this limit the source 
is as small as allowed by the uncertainty principle. The result are 
{\em vanishing} HBT radii,\cite{HSZ00} and the correlation function 
becomes completely flat (i.e. $q$-in\-dependent). [Note that in this limit 
also the residual correlation function $C_2^{\rm (res)}$ become 
$q$-independent.\cite{HSZ00}] Superficially his looks like the correlation 
function from a coherent state (``pion laser''\cite{ZC98,Pratt93}) which 
shows no Bose-Einstein correlations. This impression is, however, quite 
misleading: The correlation function still has the structure (\ref{62}), 
with an exchange term $R_2\equiv 1$ which is as large as the direct term 
(i.e. the source is still fully chaotic). Note that this limit can be 
taken while keeping the pion multiplicity distribution fixed, i.e. the 
latter does not automatically become a Bose-Einstein distribution when 
$v\to 1$. 

\section{Epilogue}\label{sec5}

Does all this matter for our real lives as heavy-ion physicists? Most 
likely not. An analysis of average pion phase-space densities at 
freeze-out from heavy-ion collisions at AGS and SPS energies\cite{Fetal} 
yields universally low values for all considered collision systems and 
beam energies. First results from RHIC appear to confirm this even at 
the much higher collision energy of $\sqrt{s}=130\,A$\,GeV.\cite{STAR} 
Apparently pions stop interacting (freeze out) only after the phase-space 
density has dropped below a critical value which is low enough for
multi-boson symmetrization effects to remain unimportant. This is good 
news: it means that we can continue to use the standard HBT 
formalism\cite{reviews} to extract source sizes from two-pion correlation 
data. It would be good, though, to confirm this by directly checking
for (the absence of) residual correlations in a high-statistics analysis
of two-particle correlation data from heavy-ion collisions. This can be
done by selecting heavy-ion collisions producing an (approximately) fixed 
number of, say, $\pi^+$ mesons and searching for a minimum at non-zero 
$q$ in the (Coulomb corrected) $\pi^+\pi^+$ correlation function.\cite{HSZ00} 
At the same time, it may be interesting to explore other fields of 
application for the elegant multi-boson symmetrization techniques 
presented here. 

\vskip 4mm
\noindent{\bf Acknowledgement:} I thank my collaborators P. Scotto and 
Q.H. Zhang for their contributions to this work,\cite{HSZ00} and the 
organizers of ISMD2000 and in particular P. L\'evai for their hospitality 
during this stimulating workshop.



\begin{thebibliography}{99}
\bibitem{fn1}
  For $N$ source currents each emitting $n_0$ pions, the normalization
  of the $n$-pion state is $\bar n^{n/2}$ with $\bar n[N] \geq n_0 N$.
  Equality holds for very dilute sources; for high phase-space densities
  $\bar n$ is much larger than $n_0 N$. This can be consistently 
  interpreted in terms of ``stimulated emission''.\cite{ZC98}
\bibitem{fn2}
  In these special cases, for certain parameter values the multiplicity 
  distribution can become unnormalizable if $v$ drops below a critical 
  value $v_{\rm c}$, due to the onset of Bose condensation.\cite{ZC98,ZSH98}
\end{thebibliography}
\end{document}